\documentclass[twocolumn]{revtex4} 
								\usepackage{amsmath}
								\usepackage{bm}
								\usepackage{fancybox}
								\usepackage[dvips]{graphicx}
								\usepackage{color}
								\usepackage{ulem}
								\usepackage{amssymb}
					\usepackage{fancyhdr}
					\renewcommand{\headrulewidth}{0pt}
\begin{document}
\title{Counter Chemotactic Flow in Quasi-One-Dimensional Path}

\author{Masashi Fujii, Akinori Awazu, and Hiraku Nishimori}
\affiliation{Department of Mathematical and Life Sciences, Hiroshima University,\\Kagami-yama 1-3-1, Higashi-Hiroshima 739-8526, Japan}

\begin{abstract}
Quasi-one-dimensional bidirectional particle flow including the effect of chemotaxis is investigated through a modification of the John-Schadschneider-Chowdhury-Nishinari model.
Specifically, we permit multiple lanes to be shared by both directionally traveling particles.
The relation between particle density and flux is studied for several evaporation rates of pheromone, and the following results are obtained: $i)$ in the low-particle-density range, the flux is enlarged by pheromone if the pheromone evaporation rate is sufficiently low, $ii)$ in the high particle-density range, the flux is largest at a reasonably high evaporation rate and, $iii)$ if the evaporation rate is at the level intermediate between the above two cases, the flux is kept small in the entire range of particle densities. The mechanism of these behaviors is investigated by observing the spatial-temporal evolution of particles and the average cluster size in the system.
\end{abstract}


\maketitle

Recently, models of various particle flows such as traffic flow, granular flow, and pedestrian flow have been investigated\cite{traffic_review_1, traffic_review_2, hysteresis_drive_model, pedestrian_model, pedestrian_counter_flow, multi_species_model, two_lane_traffic, rule_184_++C}.
In general, these systems often experience a `jammed state' with an increase in element density.
On the other hand, a population of ants taking on chemotaxis with pheromone seems to be able to avoid such state even when their density is high\cite{biATM, phaseATM, ant_pedestrian, entangled_cluster, ant_foraging}.
Thus, such biological aspects seem to provide some hints to eliminate jam in particle flows.

In this study, we investigate the possible effects of the interaction among particles through pheromone on particle flow.
We consider a model of bidirectional (i.e., counter) particle flow in which the particle deposit pheromone and determine their motions according to the pheromone field.
The model is given as a modification of a cell model of the bidirectional ant traffic introduced by John et al.\cite{biATM} (hereafter, the JSCN model).
Practically, to realize a wider class of flows rather than restricting what to traffic flow, we simply extend the JSCN model as follows:
I) The system consists of two lanes where the particles traveling in both directions share both 
lanes and are permitted to move within a lane and to shift to the other, namely, a particle can swerve on encountering the oppositely traveling another particle. 
II) The particles momentarily can go backward contrary to their traveling direction according to the rule described below, where the traveling direction is defined as the direction of destination cells at the right (or left) edge of the system.

Details of the model are given as follows.
The system consists of two lanes and each lane involves $L$ cells. 
From a cell at the left edge to that at the right edge in each lane, we give the cell index, 0, 1, $\cdots$, $L-1$.
To realize the excluded volume effect, only one particle is permitted to be contained per cell.
The particle deposits pheromone on its occupying cell.  The number of pheromone is given zero or one which increases with the secretion and decreases with evaporation.

\begin{figure}[ht]
	\begin{center}
		\includegraphics[width=72mm]{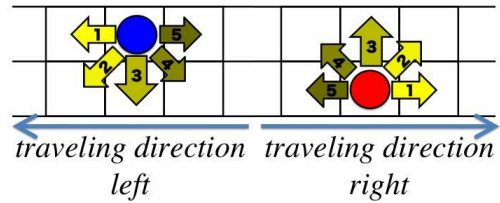}
		\caption{(Color online)
			Ordering for the selection of the moving direction of a particle at each sub-step.
		}
		\label{fig:moving direction}
	\end{center}
\end{figure}

In each sub step, one cell is randomly selected among $2 \times L$ cells.
If no particle is found but a pheromone is found in the selected cell, the pheromone in that cell is removed with a probability $f$, whose quantity is defined as the evaporation rate.
On the other hand, if the selected cell is occupied by a particle, one candidate cell for its next position is determined among neighboring unoccupied cells.
The selection among these neighboring cells is given in the following designations order:
$1)$ forward cell (traveling directional cell),
$2)$ diagonally forward cell,
$3)$ side cell,
$4)$ diagonally backward cell,
and $5)$ backward cell, as shown in Fig. \ref{fig:moving direction}.
After the candidate cell for the next position is determined, it is checked if a pheromone is left or not (one or zero) in the cell.
If a pheromone is left, the particle moves with the probability $Q$, and in the opposite case, with the probability $q$ where $Q>q$ to reflect chemotaxis.
After moving in this step, a pheromone is put in its newly located cell, and a pheromone in the leaving cell is removed with the probability $f$. 
When the particle reaches the destination cells of the right (or left) edge of the system, it changing its traveling direction to the opposite\cite{boundary}.
The same sub steps are repeated $2\times L$ times.
All of the above processes constitutes the unit time step. 

In the following, we show typical behaviors of this model.
As the initial condition, $N$ particles are set randomly in the field.
The parameters are $Q=0.95$, $q=0.25$, and $L=500$. 

\begin{figure}[ht]
	\begin{center}
		\includegraphics[width=74mm]{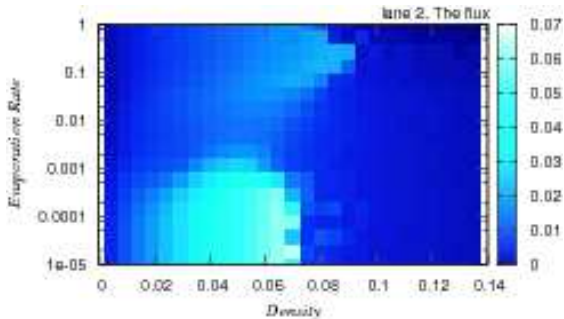}
		\caption{
			(Color online)
			Relationship among density (horizontal axis), evaporation rate of pheromone (vertical axis) and flux. A brighter tone means a higher flux.
		}
		\label{fig:bi-directional flow 1}
	\end{center}
\end{figure}

\begin{figure}
	\begin{center}
		\includegraphics[width=70mm, height=40mm]{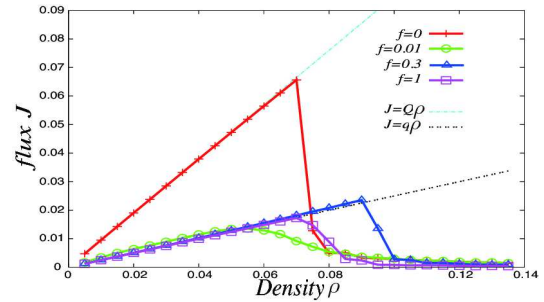}
		\caption{
			(Color online)
			$J-\rho$ relationship; for $f=0 (+)$, $0.01 (\bigcirc)$, $0.3 (\triangle)$ and $1.0 (\square)$.
		}
		\label{fig:bi-directional flow 2}
	\end{center}
\end{figure}

First, we determine the relationship between the particle density $\rho = N/(2L)$, the evaporation rate of the pheromone $f$, and the flux $J$.
Here, $J$ is given by the long-term average (averaged over time steps $0 \sim 10^7$) of the number of particles reaching the destination cells per unit time step.
Figure \ref{fig:bi-directional flow 1} shows the relationship among them.
As roughly guessed from this figure, several characteristic types of fundamental diagram are realized in different ranges of evaporation rates.
Here, specific values, namely, $f=0$, $f=0.01$, $f=0.3$, and $f=1.0$, are chosen and the fundamental diagrams realized at each respective $f$ are shown in Fig. \ref{fig:bi-directional flow 2}.

For $f =0$, $J$ linearly increases with $\rho$ as $J=Q\rho$ within the density region $\rho < 0.07$, which reflects the acceleration by a pheromone through the relation $Q>q$.
However, $J$ falls below the linear relation over a critical density $\rho=0.07$, which is much smaller than the lowest jam density given in the JSCN model because of the lane, sharing rule by both directionally traveling particles that makes jam more tenacious.

At higher evaporation rates, $f =0.3$ and $f = 1$, in the low-density region, $J$ linearly increases with $J=q\rho$, which means flux is kept less than that in the above case of $f =0$.
In the case of $f=0.3$, however, the linearly increasing region extends up to $\rho=0.09$, this is, larger flux than the case of $f=0$ is realized in a finite range beyond $\rho = 0.07$ as seen in Fig. 3.
In this way, in a certain density range beyond $\rho=0.07$, a larger flux is realized with a reasonably higher evaporation rate than that in the no-evaporation case.
 
At an intermediate evaporation rate, $f=0.01$, $J$ is barely larger than $J=q\rho$ in a small-$\rho$ region.
At approximately $\rho=0.055$, $J$ begins to decrease, but the shift from the increasing region to the decreasing region is not so sharp.
In this case, $J$ is kept low throughout the entire range of densities.


Next, to understand the mechanism giving these characteristic fundamental diagrams, we determine the spatiotemporal evolution of particles and the average cluster size for each situation.
To this purpose, we define the term `cluster', here, as a group of more than one particles, which is inseparable by more than one contiguous empty columns.
Here, a `column' is a set of cells having identical cell indexes.
Furthermore, `entangled cluster' is defined as a cluster consisting of oppositely traveling particles that stays at almost the same location.

\begin{figure}[ht]
	\begin{center}
		\includegraphics[width=70mm, height=40mm]{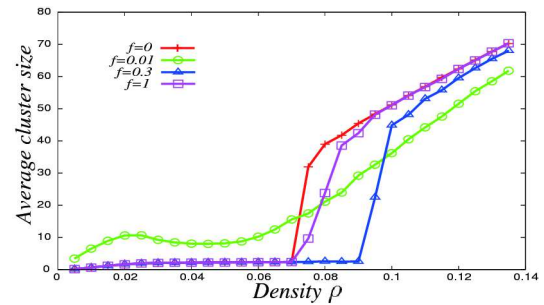}
		\caption{
			(Color online)
			Average cluster size as a function of $\rho$, for the evaporation rate: $f=0 (+)$, $0.01 (\bigcirc)$, $0.3 (\triangle)$ and $1.0 (\square)$.
		}
		\label{fig:cluster}
	\end{center}
\end{figure}

First, we see the cases of no evaporation, $f=0$.
At a density $\rho = 0.04$, particles move without forming large clusters, as shown in Figs. \ref{fig:cluster} and \ref{fig:f=0,d=0.04}, and each particle moves at a velocity close to $Q$.
In Fig. \ref{fig:f=0,d=0.04}, the red/blue dots indicate that only right/left directionally traveling particle(s) is (are) located in the corresponding columns, and green dots indicate that both directional particles are contained in those columns.
Figures \ref{fig:f=0,d=0.10}-\ref{fig:f=0.01,d=0.04,d=0.07} are plotted in the same manner.
As seen in Fig. 5, when two oppositely traveling particles or small clusters encounter, they smoothly weave through the other particle or cluster and, accordingly, avoid jam formation.
At a higher density $\rho=0.1$, through the encounter of two oppositely traveling clusters larger than those in the above case [Fig. \ref{fig:cluster}], they easily form an entangled cluster. 
The size of this entangle cluster grows with time through frequent encounters with other incoming clusters [Fig. \ref{fig:f=0,d=0.10}(a)], and this cluster finally absorbs most of the particles in the system, thereby blocking flow [Fig. \ref{fig:f=0,d=0.10}(b)].
Then $J$ drastically decreases to almost zero [Fig. \ref{fig:bi-directional flow 2}].
Between these two densities, at $\rho=0.075$, the system alternately stays in jammed and smoothly flowing states [Fig. \ref{fig:f=0,d=0.075}(a)].
Then, the average flow $J$ takes an intermediate value.

The above result is unlike the dynamics as reported in the JSCN model in which the jammed state containing most particles in the system 
is not dissolved once formed.
This characteristic dynamics in the present model is realized because particles are permitted to move against the traveling direction according to their temporal situations.
By combining this backward motion of particle with the stochastic selection of a cells performing each sub step, the system behaves like a double-well potential system in which a particle goes back and forth between two stable states, though they are not exactly the same.

\begin{figure}
	\begin{center}
		\includegraphics[width=72 mm]{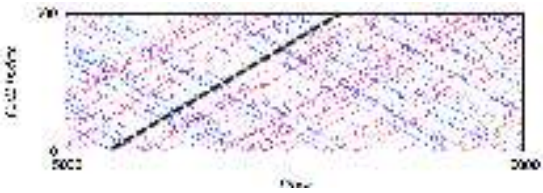}
		\caption{
			(Color online)
			Typical spatiotemporal behaviors of particles for $\rho = 0.04$ and $f=0$.
			The broken line indicates the inclination $Q$ meaning the trajectory of free particles in a field filled with a pheromone.
		}
		\label{fig:f=0,d=0.04}
	\end{center}
\end{figure}
\begin{figure}
	\begin{center}
		(a)
			\hspace{-2.8mm}
			\begin{minipage}[t][][b]{72 mm}
				\includegraphics[width=72 mm]{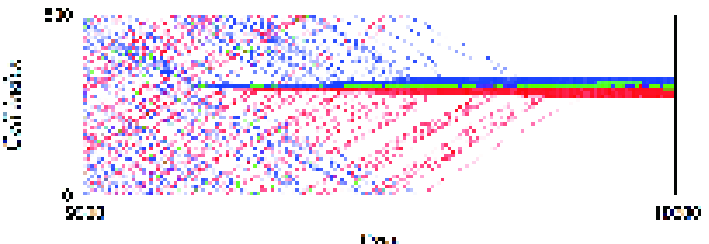}
			\end{minipage}
			\hspace{-1mm}
		(b)
			\hspace{-2.8mm}
			\begin{minipage}[t][][b]{72mm}
				\includegraphics[width=72mm]{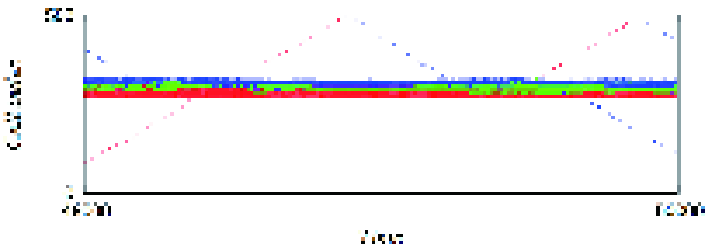}
			\end{minipage}
		\caption{
			(Color online)
			Typical spatiotemporal behaviors of particles in
			(a) transient with the initiation of entangled cluster, and
			(b) the steady state for
			$\rho = 0.1$ and $f=0$.
		}
		\label{fig:f=0,d=0.10}
	\end{center}
\end{figure}
\begin{figure}
	\begin{center}
		(a)
			\hspace{-2.8mm}
			\begin{minipage}[t][][b]{72 mm}
				\includegraphics[width=72 mm]{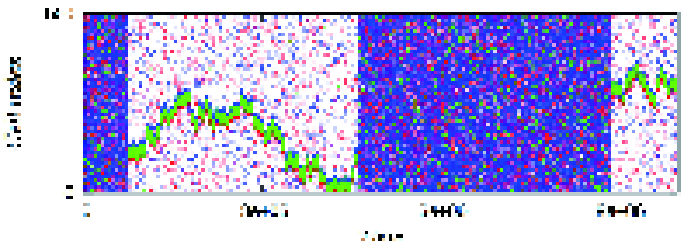}
			\end{minipage}\\[2 mm]
		(b)
			\hspace{-2.8mm}
			\begin{minipage}[t][][b]{72 mm}
				\includegraphics[width=72 mm]{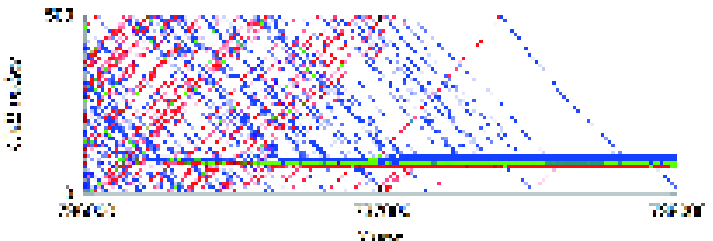}
			\end{minipage}
			\hspace{-1mm}
		(c)
			\hspace{-2.8mm}
			\begin{minipage}[t][][b]{72mm}
				\includegraphics[width=72mm]{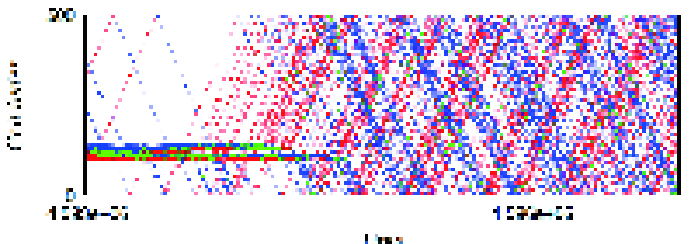}
			\end{minipage}
		\caption{
			(Color online)
			Spatial-temporal behaviors of particles:
			(a) transitions between two different typical states,
			(b) the initiation of entangled cluster, and
			(c) the termination of entangled cluster,
			where (b) and (c) are picked up from (a) for $\rho = 0.075$ and $f=0$.
		}
		\label{fig:f=0,d=0.075}
	\end{center}
\end{figure}

Below, we discuss the cases of a high evaporation rate $f=0.3$.
At a density $\rho= 0.06$, particles behave similarly to those in the case of $f=0$ and $\rho=0.04$.
Namely, they move without making large clusters, as seen in Fig. \ref{fig:cluster}.
Each particle moves at an average velocity close to $q$ [Fig. \ref{fig:f=0.3,d=0.06,d=0.12} (a)] which reflects the high evaporation rate of a pheromone.
When two oppositely traveling particles (or small clusters) encounter, they smoothly weave through each other to avoid jam formation.
As shown in Fig. \ref{fig:cluster}, at this evaporation rate, the clusters are kept small until $\rho=0.09$, accompanied by a smoothly flowing state and a linearly increasing part in the fundamental diagram in Fig. \ref{fig:bi-directional flow 2}.
The reason for the smooth flow being sustained up to a higher density than that in the case of $f=0$ is yet to be studied.
At a high density $\rho=0.12$, particles behave similarly to those in the case of $f=0$ and $\rho=0.1$.
Namely, through the encounter of oppositely traveling two clusters, they easily form an entangled cluster that grows with time through frequent encounters with other incoming of other clusters, thereby finally absorbing most of particles in the system to block flow.
Therefore, $J$ becomes almost zero [Fig. \ref{fig:f=0.3,d=0.06,d=0.12}(b)].
Between these two densities, at $\rho=0.1$, the system alternately takes one of the above mentioned two states.

\begin{figure}
	\begin{center}
		(a)
			\hspace{-2.8mm}
			\begin{minipage}[t][][b]{72mm}
				\includegraphics[width=72mm]{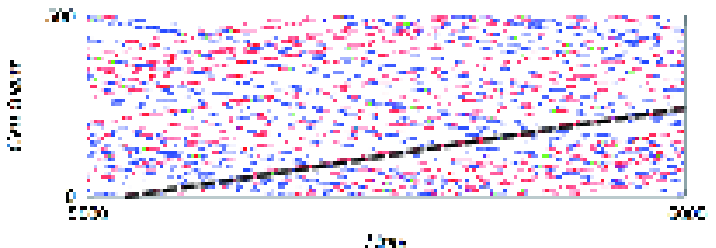}
			\end{minipage}
			\hspace{-1mm}
		(b)
			\hspace{-2.8mm}
			\begin{minipage}[t][][b]{72mm}
			\includegraphics[width=72mm]{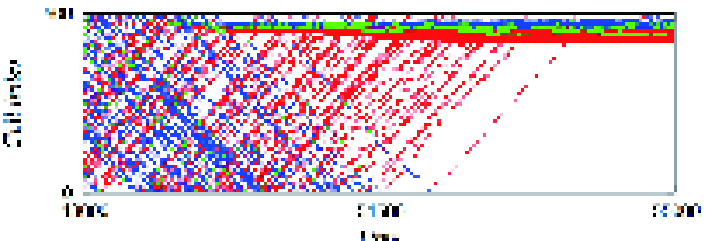}
			\end{minipage}
		\caption{
			(Color online)
			Typical spatiotemporal behaviors of particles for (a) $\rho = 0.06$ and (b) $\rho = 0.12$ with $f=0.3$.
			The broken line indicates the inclination $q$, which is the trajectory of free particles in a field without a pheromone.
		}
		\label{fig:f=0.3,d=0.06,d=0.12}
	\end{center}
\end{figure}

Then, we consider the case of an intermediate evaporation rate, $f=0.01$.
At $\rho=0.04$, around which $J$ almost linearly increases with $\rho$, particles move as medium-sized clusters (Fig. \ref{fig:cluster}) with a velocity between $q$ and $Q$.
When two oppositely traveling clusters encounter, they temporally form an entangled cluster; however, it is immediately resolved without growing to include most particles in the system.
Before long, a new entangled cluster is formed in the system to repeat the same process.
At a critical density $\rho=0.055$, $J$ decreases.
However, unlike those in the cases of $f=0.0$ and $f=0.3$, the spatiotemporal behavior of the system does not markedly change at the critical density, nor the average cluster size discontinuously changes.
As seen in Fig. \ref{fig:f=0.01,d=0.04,d=0.07}, even beyond the critical density, entangled clusters are not sustained as in the cases of $f=0$ and $f=0.3$; instead, their formation and dissolution are repeated in the system, although the average cluster size gradually increases with density.
Hence, flux is kept  low but not zero, as seen in Fig. \ref{fig:bi-directional flow 2}.

\begin{figure}
	\begin{center}
		(a)
		\hspace{-2.8mm}
		\begin{minipage}[t][][b]{72mm}
			\includegraphics[width=72mm]{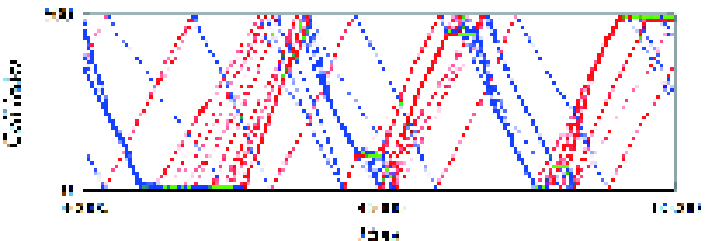}
		\end{minipage}
		\hspace{-1mm}
		(b)
		\hspace{-2.8mm}
		\begin{minipage}[t][][b]{72mm}
			\includegraphics[width=72mm]{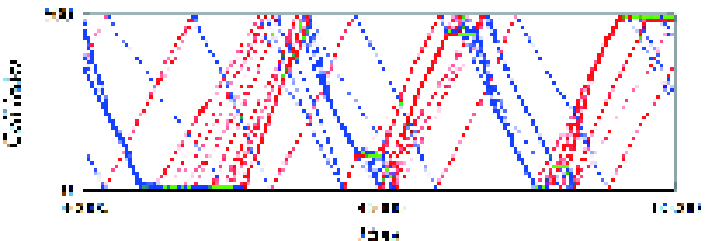}
		\end{minipage}
		\caption{
			(Color online)
			Typical spatiotemporal behaviors of particles for
			(a) $\rho = 0.04$ and $f=0.01$ and
			(b) $\rho = 0.07$ and $f=0.01$.
		}
		\label{fig:f=0.01,d=0.04,d=0.07}
		\end{center}
\end{figure}


In this study, the dynamical and statistical aspects of two lanes bi-directional particle flow including the effect of chemotaxis were investigated.
We found the following relationships among particle density, the evaporation rate of pheromones, and flux:
I) In the low-density region, the flux of particles is largest if the evaporation rate is sufficiently low;
II) in the higher-density region, a reasonably high evaporation rate enhances flux; and
III) flux is kept small at all density region if evaporation rate is intermediate between those in cases I and II.
The analysis of the balance (or the breakdown of the balance) between the growth rate and dissolution rate of entangle clusters seems to be the key to further investigations, as has already been mentioned in a previous study by Kunwar et al.\cite{entangled_cluster}

The authors are grateful to R. Kawai, M. Akiyama and members of the Mathematical Society of Traffic Flow for useful discussion and information.
This study was supported in part by The Global COE Program G14 (Formation and Development of Mathematical Sciences Based on Modeling and Analysis) of the Ministry of Education, Culture, Sports, Science and Technology of Japan.

\vspace{-8mm}
\begin{center}
\rule[0mm]{80mm}{0.05mm}
\end{center}
\vspace{-10mm}



\begin{thebibliography}{99}
\bibitem{boundary} If this boundary condition is changed to periodic boundary conditions, the following results are kept qualitatively the same.
\end{thebibliography}
\end{document}